\documentclass[aps,prl,twocolumn,superscriptaddress,longbibliography]{revtex4-1}

\usepackage{amsfonts}
\usepackage{amsmath,amssymb}
\usepackage{units}
\usepackage{graphicx}

\begin{document}


\title{$\mbox{Bi}_{1}\mbox{Te}_{1}$: a dual topological insulator}


	\author{Markus Eschbach} 		\altaffiliation{These authors contributed equally to this work.}
									\affiliation{Peter Gr\"unberg Institute and JARA-FIT, Forschungszentrum J\"ulich GmbH, 52425 J\"ulich, Germany}
	\author{Martin Lanius}			\altaffiliation{These authors contributed equally to this work.}
									\affiliation{Peter Gr\"unberg Institute and JARA-FIT, Forschungszentrum J\"ulich GmbH, 52425 J\"ulich, Germany}		
	\author{Chengwang Niu}			\altaffiliation{These authors contributed equally to this work.}
									\affiliation{Peter Gr\"unberg Institute and JARA-FIT, Forschungszentrum J\"ulich GmbH, 52425 J\"ulich, Germany}			
	\author{Ewa M\l y\'{n}czak} 	
									\affiliation{Peter Gr\"unberg Institute and JARA-FIT, Forschungszentrum J\"ulich GmbH, 52425 J\"ulich, Germany}
									\affiliation{Faculty of Physics and Applied Computer Science, AGH University of Science and Technology, al. Mickiewicza 30, 30-059 Krakow, Poland}
	\author{Pika Gospodari\v{c}} 	
									\affiliation{Peter Gr\"unberg Institute and JARA-FIT, Forschungszentrum J\"ulich GmbH, 52425 J\"ulich, Germany}
	\author{Jens Kellner}			\affiliation{II. Institute of Physics B and JARA-FIT, RWTH Aachen University, 52074 Aachen, Germany}
	\author{Peter Sch\"uffelgen}	
									\affiliation{Peter Gr\"unberg Institute and JARA-FIT, Forschungszentrum J\"ulich GmbH, 52425 J\"ulich, Germany}	
	\author{Mathias Gehlmann}		
	\author{Sven D\"oring}			
	\author{Elmar Neumann}			
									\affiliation{Peter Gr\"unberg Institute and JARA-FIT, Forschungszentrum J\"ulich GmbH, 52425 J\"ulich, Germany}
	\author{Martina Luysberg}		\affiliation{Peter Gr\"unberg Institute and Ernst Ruska-Centre for Microscopy and Spectroscopy with Electrons, Forschungszentrum J\"ulich GmbH, 52425 J\"ulich, Germany}
									
	\author{Gregor Mussler}			
									\affiliation{Peter Gr\"unberg Institute and JARA-FIT, Forschungszentrum J\"ulich GmbH, 52425 J\"ulich, Germany}	
	\author{Lukasz Plucinski}		\email[Author to whom correspondence should be addressed: ]{l.plucinski@fz-juelich.de}
									\affiliation{Peter Gr\"unberg Institute and JARA-FIT, Forschungszentrum J\"ulich GmbH, 52425 J\"ulich, Germany}	
	\author{Markus Morgenstern}		\affiliation{II. Institute of Physics B and JARA-FIT, RWTH Aachen University, 52074 Aachen, Germany}
	\author{Detlev Gr\"utzmacher}	
									\affiliation{Peter Gr\"unberg Institute and JARA-FIT, Forschungszentrum J\"ulich GmbH, 52425 J\"ulich, Germany}	
	\author{Gustav Bihlmayer}		
									\affiliation{Peter Gr\"unberg Institute and JARA-FIT, Forschungszentrum J\"ulich GmbH, 52425 J\"ulich, Germany}	
	\author{Stefan Bl\"ugel}		
									\affiliation{Peter Gr\"unberg Institute and JARA-FIT, Forschungszentrum J\"ulich GmbH, 52425 J\"ulich, Germany}	
	\author{Claus M. Schneider}		
									\affiliation{Peter Gr\"unberg Institute and JARA-FIT, Forschungszentrum J\"ulich GmbH, 52425 J\"ulich, Germany}	

\date{\today}

\begin{abstract}
A combined theoretical and experimental study reveals evidence for
the dual topological insulating character of the stoichiometric natural
superlattice phase $\mathrm{Bi_{1}Te_{1}}=\mathrm{[Bi_{2}]_{1}[Bi_{2}Te_{3}]_{2}}$,
being a stack of alternating Bi bilayers and two quintuple layers
of $\mathrm{Bi_{2}Te_{3}}$. We identify $\mathrm{Bi_{1}Te_{1}}$
by density functional theory to exhibit a non trivial time-reversal
symmetry-driven character of $\mathbb{Z}_{2}=(0;001)$
and additionally a mirror-symmetry induced mirror Chern number of
$n_{{\cal M}}=-2$, which indicates that $\mathrm{Bi_{1}Te_{1}}$
is both a weak topological insulator and a topological crystalline
insulator. The coexistence of the two phenomena preordains distinct
crystal planes to host topological surface states that are protected
by the respective symmetries. The surface perpendicular to the stacking
direction is the 'dark' surface of the weak topological insulator,
while hosting mirror-symmetry protected surface states along the $\overline{\Gamma\mathrm{M}}$
direction at non-time-reversal invariant momenta points. We confirm
the stacking sequence of our MBE-grown $\mathrm{Bi_{1}Te_{1}}$ thin
films by X-ray diffraction and transmission electron microscopy, and
find indications of the topological crystalline and weak topological
character in the surface electronic spin structure by spin- and angle-resolved
photoemission spectroscopy, which nicely match the results from density
functional theory. 
\end{abstract}


\maketitle

\section{Introduction}

Topological insulators (TIs) are bulk insulating materials which exhibit
perfect metallic conductivity on their boundary via electronic edge
(in 2D TIs) or surface states (in 3D TIs) that are guaranteed by the
topological character of the bulk band structure \cite{Hasan2010,Ando2013}.
Electrons in these boundary states are highly spin-polarized and their
spin and momentum is locked to each other by spin-orbit coupling,
creating helical spin textures and making TIs highly attractive for
spintronic applications \cite{Hsieh2009b}. One of the most exciting
aspects of 3D TIs is the fact that their surface inevitably hosts
these metallic surface states as long as the symmetry defining the
topological character is not destroyed \cite{FuKane2007,Murakami:07.1}.
In a strong topological insulator (STI), for instance, time-reversal
symmetry protects these states on all surfaces. Weak topological insulators
(WTIs), on the other hand, display protected metallicity only at surfaces
with a certain orientation, while other surfaces remain insulating.
The latter can be understood in a simple picture, where a stack of
two-dimensional TIs forms a WTI with metallic edge states inherited
from the 2D TI but with an insulating surface plane (the 'dark side'')
normal to the stacking direction. Finally, in topological crystalline
insulators (TCIs), where the symmetry with respect to a mirror plane
defines the topology, the metallic surface states can be found on
surfaces perpendicular to these mirror planes \cite{Teo:08.1,Fu:11.1}.\\
For the case of $\mathrm{Bi_{2}Te_{3}}$ it was already shown that
a single compound can belong both to the class of strong TIs and TCIs
\cite{Rauch:14.1}. It is intriguing to think of the possibilities
opened by symmetry breaking that can destroy certain surface states
while keeping others intact. For example one could imagine a material
that is both a WTI and a TCI and has all surfaces covered with metallic
edge states, where the mirror plane of the TCI is normal to the dark
side of the WTI. Then, a magnetic field would destroy the topological
protection of the states originating from the topological properties
that define the WTI, while the mirror-symmetry protected states remain
intact; Likewise, small structural distortions can destroy the mirror
plane without affecting the edge states arising from time-reversal
symmetry.\\
In the search for such a material, we start from $\mathrm{Bi_{2}Te_{3}}$
with the above mentioned properties and from the Bi bilayer that is
known to be a 2D TI \cite{Murakami2006,Wada2011,Yang2012}. It is
possible to produce natural superlattices $[\mathrm{Bi}_{2}]_{x}[\mathrm{Bi}_{2}\mathrm{Te}_{3}]_{y}$
from hexagonal, metallic Bi bilayer (BLs) and semiconducting $\mathrm{Bi_{2}Te_{3}}$
quintuple layer (QLs) building blocks in a wide range of $x$ and
$y$ \cite{Bos2007,Bos:12.1,Cava2013,Isaeva2013}. While $\mathrm{Bi_{2}Te_{3}}$
consists of only QL building blocks, the unit cell of $\mathrm{Bi_{1}Te_{1}}$
phase is made out of a stacking sequence of a single BL interleaved
with two subsequent QLs. The size of the unit cell along the stacking
direction, i.e. the $c$ lattice constant, varies quite severely among
the different stable compounds, which makes them easily distinguishable
in a diffraction experiment. Recently, $\mathrm{Bi_{4}Se_{3}}$ (i.e.
$x=y=1$) was investigated in some detail and characterized as topological
semimetal \cite{Valla2012}. For practical applications, however,
an insulating bulk material is preferable. \\
\\
In this article, we identify the stoichiometric natural superlattice
$\mathrm{Bi_{1}Te_{1}}$ (i.e. $x=1,\,y=2$) as a semiconductor with
a small bandgap of about $\unit[0.1]{eV}$ with the desired properties:
$\mathrm{Bi_{1}Te_{1}}$ is both a WTI and a TCI where the combination
of both properties leads to topologically protected surface states
(TSS) on all sides of the crystal. Our density functional theory (DFT)
calculations predict a $\mathbb{Z}_{2}$ class of $(0;001)$ and a
mirror Chern number $n_{{\cal M}}=-2$. We find two characteristic
surface states of the TCI on the (0001) surface, regardless of the
surface termination. A very similar situation has been reported recently
for $\mathrm{Bi_{2}TeI}$, theoretically \cite{Rusinov2016}. \\
We demonstrate that $\mathrm{Bi_{1}Te_{1}}$ can be grown in form
of high quality thin films on Si(111) by molecular beam epitaxy (MBE).
Its layered structure is confirmed by scanning transmission electron
microscopy (STEM) and X-ray diffraction (XRD), confirming a repeated
stacking sequence of 2QL of $\mathrm{Bi_{2}Te_{3}}$ and a single
Bi BL. We investigate the electronic structure of $\mathrm{Bi_{1}Te_{1}}$
by means of spin- and angle-resolved photoemission (spin-ARPES). The
experimental results are compared to the well known prototypical 3D
STI $\mathrm{Bi_{2}Te_{3}}$ ($x=0$, $y=3$). Spin-ARPES reveals
that in $\mathrm{Bi_{1}Te_{1}}$ the surface states close to $E_{\mathrm{F}}$
exhibit a nearly vanishing spin polarization in contrast to the time-reversal
symmetry driven TSS in $\mathrm{Bi_{2}Te_{3}}$. Furthermore, our
spectra taken along non-high-symmetry lines reveal band crossings
at non-time-reversal invariant momenta (TRIM) points that can be associated
with surface states protected by mirror symmetry and the TCI character
of $\mathrm{Bi_{1}Te_{1}}$.

\section{Results \& Discussion}

\subsection{Ab-initio calculations. }

\begin{figure*}
	\centering
	\includegraphics[scale=0.45]{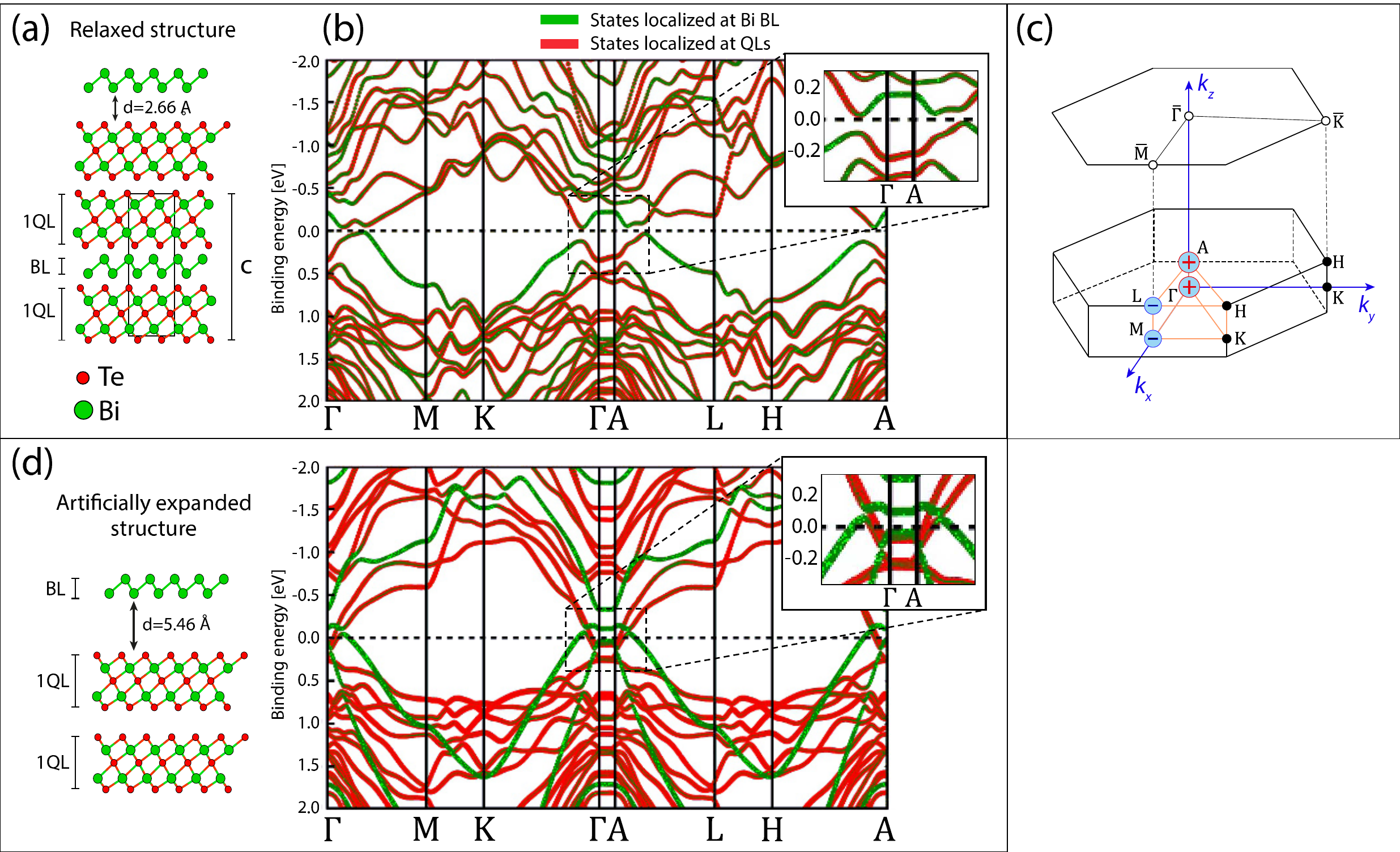}
	\caption{\label{fig:bulk} Bulk band structures of $\mathrm{Bi_{1}Te_{1}}$.
	(a) Simple sketch of the crystal structure of $\mathrm{Bi_{1}Te_{1}}$.
	The unit cell consists of 1 Bi BL and 2QLs. (b) and (d) show the bulk
	band structure calculation in the structurally relaxed geometry and
	with artificially expanded distances between the BL and the QLs, respectively.
	States localized mostly in the BL are marked in green, while the states
	localized mostly in the 2QLs are shown in red. In (b), the band structure
	of the BL shows an inverted gap about $\unit[0.2]{eV}$ above the
	Fermi level ($E_{{\rm F}}$ marked with a dashed line). (c) Bulk and
	surface Brillouin zone with parity product of the TRIM points resulting
	in +1 (red '+') or -1 (blue '-'). $k_{z}$ direction corresponds to
	the stacking direction. $\Gamma\mathrm{AML}$, i.e. $(k_{x},k_{z})$-plane,
	marks a mirror plane. }
\end{figure*}

Figure \ref{fig:bulk}(a) depicts a schematic model of the crystal
structure of $\mathrm{Bi_{1}Te_{1}}$, indicating Te (green) and Bi
(orange) atoms as well as one unit cell defined by the lattice constant
$c$ along the stacking direction. The separation of the layered structure
into QLs and Bi BLs is marked. The bulk band structure of $\mathrm{Bi_{1}Te_{1}}$
in the relaxed structural geometry is presented in Fig. \ref{fig:bulk}(b).
Spin-orbit coupling is included in this calculation and the color
represents the localization of the electronic states at the BL (green)
or at the QL (red). As one can see, there are no states at the Fermi
level $E_{\mbox{F}}$, reflecting the insulating character with an
energy gap of $\unit[73]{meV}$. The states around the Fermi level
alternate between BL- and QL-related, where the highest occupied levels
at the time-reversal invariant momenta $\Gamma$ and A stem from QLs
(red), while the lowest unoccupied states originate from the BL (green),
and vice versa for the M and L points. Since the crystal possesses
spatial inversion symmetry, the parity of the states can be calculated
and the topological index $\mathbb{Z}_{2}$ can be deduced according
to ref. \cite{Fu2007} based on the product of the parities of all
occupied bands at the eight TRIMs, i.e., one at $\Gamma$, one at
A, three at M, and three at L. The result is shown in Fig. \ref{fig:bulk}(c)
for the corresponding TRIMs, $\Gamma$ and A have parity products
of $+1$ (red '+') while M and L have $-1$ (blue '-'), leading to
a topological invariant $\mathbb{Z}_{2}=(0;001)$. Therefore, $\mathrm{Bi_{1}Te_{1}}$
is a weak topological insulator with the (0001) surface, which is
perpendicular to the stacking direction, being the ``dark'' surface
and being free of time-reversal symmetry protected surface states.\\
\\
It is tempting to relate the WTI property to the fact that both, the
Bi bilayer and the 2QLs $\mathrm{Bi_{2}Te_{3}}$ are 2D TIs such that
the WTI results from a simple stacking of 2D TIs in the $c$-direction.
However, our band structure calculations in Fig. \ref{fig:bulk}(d),
introducing artificially expanded distances between the BL and the
QLs, show a more complex scenario: If the BL is sufficiently separated
from the 2QLs, the states can be decomposed in contributions from
the two components (green = Bi BL and red = 2QLs, respectively). But,
due to charge transfer, the inverted gap of the BL is shifted above
the Fermi level and, accordingly, some of the 2QL $\mathrm{Bi_{2}Te_{3}}$
conduction band states are below $E_{{\rm F}}$. Only the hybridization
of the BL states with the QL states opens up the gap that leads to
the insulating bulk structure in Fig. \ref{fig:bulk}(b), as can be
nicely deduced from the changing color of the bands along the $k$-directions.
Nevertheless, the topological character of the stacked film remains
non-trivial. A similar complexity is also found for the first confirmed,
stacked weak TI $\mathrm{Bi_{14}Rh_{3}I_{9}}$ \cite{Pauly2015,Rasche2013,Pauly2016}.

\begin{figure}

	\includegraphics[width=0.95\columnwidth]{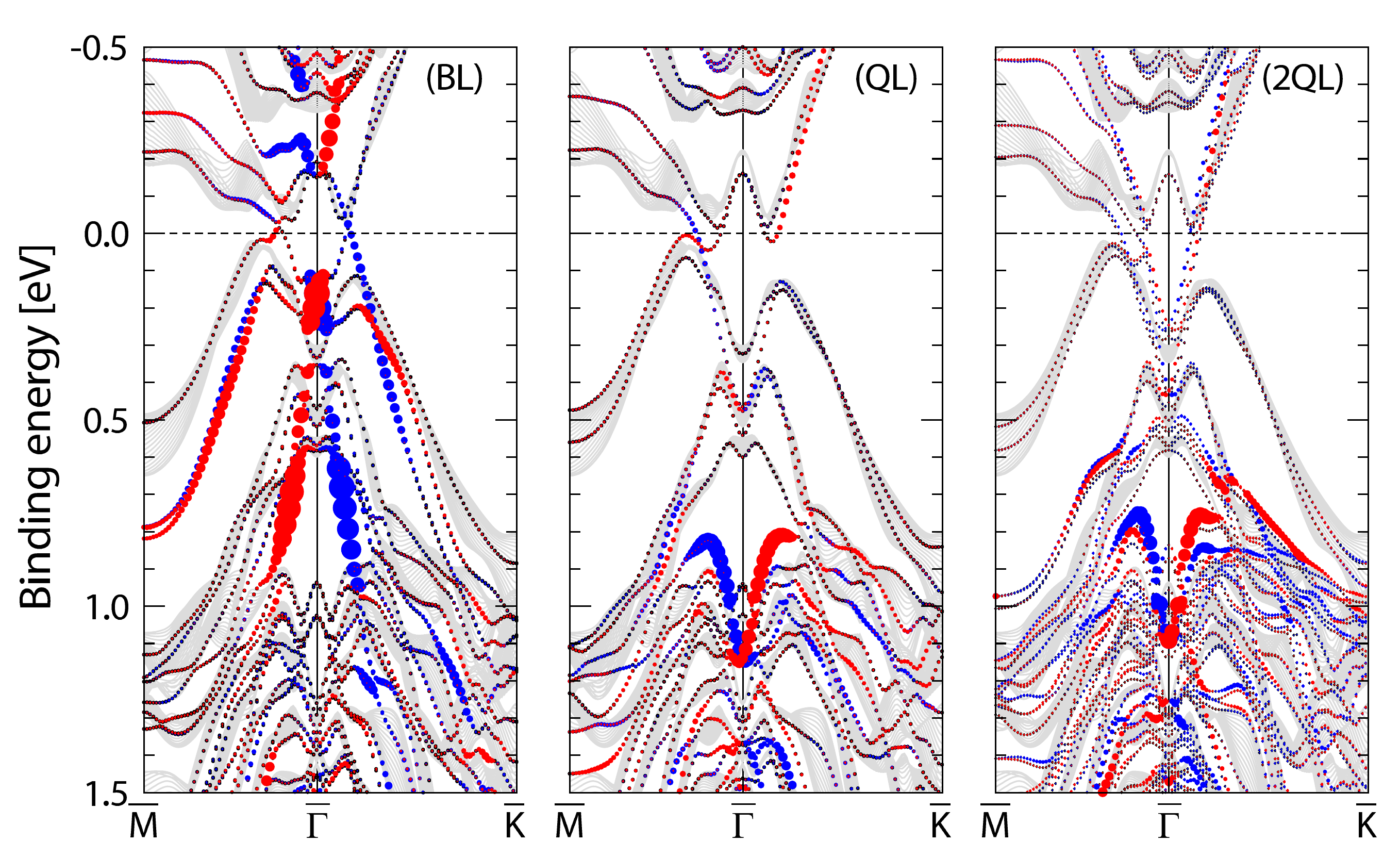} 
	\caption{\label{fig:surface DFT} Spin-resolved DFT surface electronic structure
	calculations along $\overline{\mathrm{M\Gamma K}}$ of slabs of $\mathrm{Bi_{1}Te_{1}}$
	terminated by a Bi BL (left), a single QL (center), and two QLs (right)
	with bulk projected bands in gray and surface bands in colors. The
	size of the symbols corresponds to the spin-polarization in the first
	four layers of a slab, the color (red/blue) indicates the orientation
	of the spins with respect to a direction perpendicular to the momentum
	and surface normal.}
\end{figure}

Next, we want to examine the surface bandstructure on the $(0001)$
surface, which heavily depends on the precise surface termination.
Due to the layered crystal structure and the weak van der Waals bonds
between the subsequent building blocks, there exist three natural
cleavage planes and thus surface terminations, i.e., 1Bi BL, 1QL,
and 2QL. Figure \ref{fig:surface DFT} depicts the spin-resolved surface
bandstructure for the respective terminations. In all cases an even
number of Fermi level crossings is found in $\overline{\Gamma\mathrm{M}}$
direction, while along $\overline{\Gamma\mathrm{K}}$ a bandgap between
valence- and conduction band is formed (although it can be very small).
We see that, although there are surface states, the bandstructure
is compatible with a WTI phase. Nevertheless, it is remarkable that
the bands along $\overline{\Gamma\mathrm{M}}$ show a band crossing
for all possible terminations, reminiscent of Dirac-like cones observed
in topological crystalline insulators. Additional evidence that this
crossing is protected by a mirror symmetry in the crystal comes from
the observation that the crossing is lifted when the surface atoms
are displaced in $[11\overline{2}0]$ direction, breaking this symmetry
(not shown here).\\
To check for the possibility that $\mathrm{Bi_{1}Te_{1}}$ is a TCI,
we determined the mirror Chern number of the bulk phase. In the $(k_{x},k_{z})$
plane in reciprocal space (Fig. \ref{fig:bulk}(c)) all Bloch states
can be distinguished by their eigenvalues with respect to a mirror
operation in the $(1\overline{1}00)$ plane. To calculate their corresponding
Berry phases as well as the Chern numbers, we construct a tight-binding
Hamiltonian based on the maximally localized Wannier functions. The
Chern numbers of all occupied bands for the opposite mirror eigenvalues
$+i$ and $-i$ are $n_{+i}=2$ and $n_{-i}=-2$, respectively, and
therefore the mirror Chern number $n_{{\cal M}}$ \cite{Teo:08.1},
given as $n_{{\cal M}}=(n_{-i}-n_{+i})/2$, is $n_{\mathcal{M}}=-2$,
confirming the fact that $\mathrm{Bi_{1}Te_{1}}$ is a TCI.\\
Lets discuss the individual features of the different terminations
shown in Fig. \ref{fig:surface DFT} in more detail. The BL terminated
surface is characterized by surface states that disperse from the
$\overline{\Gamma}$ point at $\unit[-0.2]{eV}$ downwards, very similar
to features observed for a single Bi bilayer on $\mathrm{Bi_{2}Te_{3}}$
\cite{Hirahara2011,Miao2013}. The steeply dispersing bands near $\overline{\Gamma}$,
crossing at $\unit[0.3]{eV}$, are a characteristic feature that is
also observed for a Bi-rich termination of $\mathrm{Bi_{4}Se_{3}}$
\cite{Gibson2013}. In contrast, the QL-terminated surface (center
panel of Fig. \ref{fig:surface DFT}) shows near the Fermi level only
the linear crossing of bands that is protected by mirror symmetry.
Strongly spin-polarized surface states are observed around $\unit[1.0]{eV}$,
which are similar to the states that also characterize the surface
of $\mathrm{Bi_{2}Te_{3}}$ \cite{Herdt2013} or $\mathrm{Sb_{2}Te_{3}}$
\cite{Pauly2012a}. These Rashba-split surface states equally appear
on the 2QL terminated surface (right in Fig. \ref{fig:surface DFT}).
In this case more states appear near the Fermi level (although leading
to an even number of crossings, compatible with the weak topological
character).\\

\subsection{Crystallographic structure.}

\begin{figure*}
	\centering
	\includegraphics[scale=0.8]{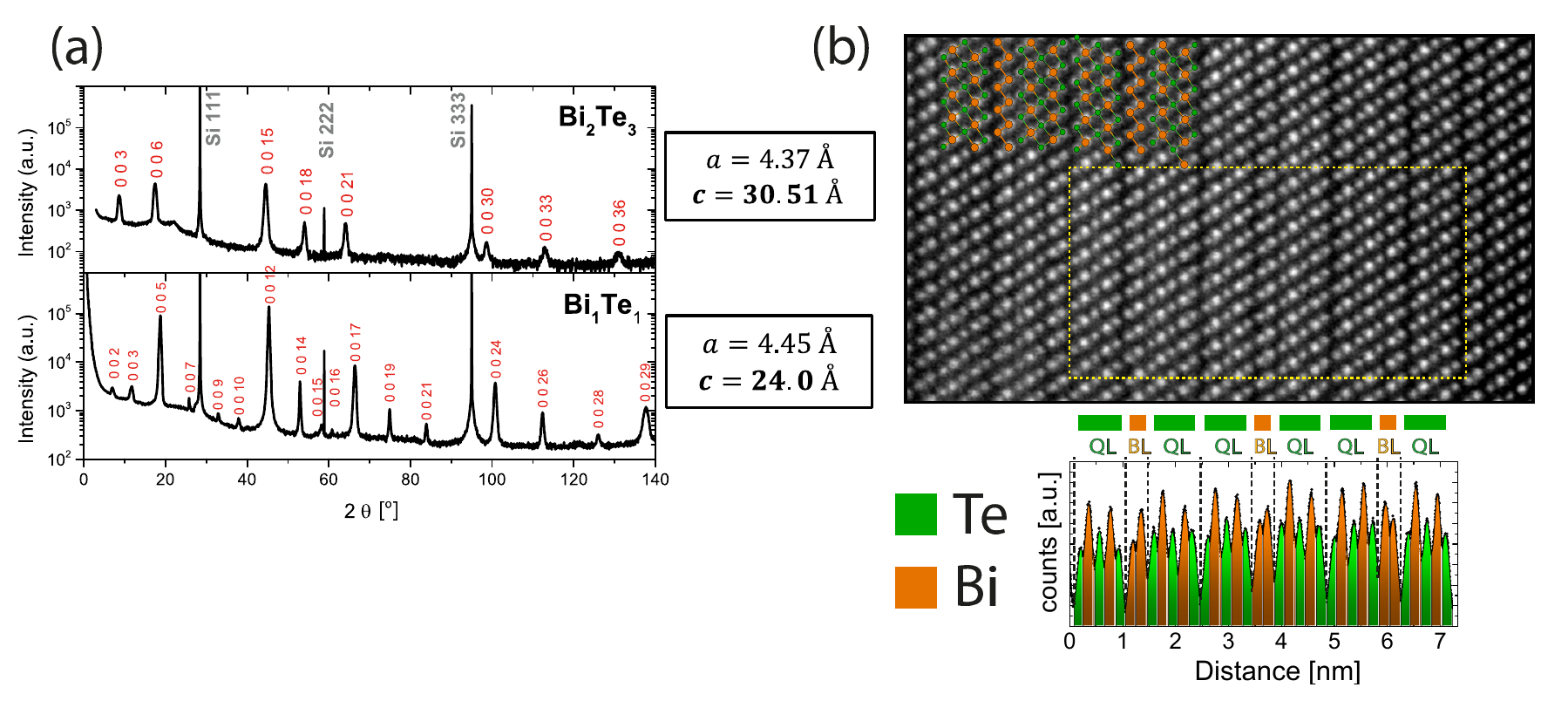}

	\caption{\label{fig:Bulk-characterization}Bulk characterization of $\mathrm{Bi_{1}Te_{1}}$
	thin films. (a) Comparative XRD $\omega/2\theta$ scans for both $\mathrm{Bi_{2}Te_{3}}$
	and $\mathrm{Bi_{1}Te_{1}}$ films averaged over the entire crystal
	with the derived $a$ and $c$ lattice constants, as highlighted.
	(b) Local STEM image of a $\unit[39]{nm}$ thick $\mathrm{Bi_{1}Te_{1}}$
	film confirming the high bulk crystalline quality. The contrast in
	the image scales with the atomic number squared $(Z^{2})$, i.e.,
	bright = Bi, darker = Te. QLs and Bi BLs separated by van der Waals
	gaps can be identified. The yellow frame marks the region over which
	the line profile below is measured while averaging in vertical direction.
	QLs and BLs are denoted and Bi and Te atoms are displayed by orange
	and green columns, respectively. }
\end{figure*}

	Figure \ref{fig:Bulk-characterization} shows the experimental characterization
	of the bulk crystal structure of our $\mathrm{Bi_{1}Te_{1}}$ thin
	films via XRD (a) and STEM (b). From the $\omega/2\theta$ scans in
	Fig. \ref{fig:Bulk-characterization}(a), the crystal phase was determined
	by comparing the peak positions with the calculated Bragg reflections,
	both for $\mathrm{Bi_{1}Te_{1}}$ and $\mathrm{Bi_{2}Te_{3}}$. The
	in-plane and out-of-plane lattice constants were measured precisely
	by defining the reciprocal space positions of the (1,0,-1,16) reflection
	for $\mathrm{Bi_{1}Te_{1}}$ and the (1,0,-1,20) reflection for $\mathrm{Bi_{2}Te_{3}}$
	(see section I in the supplementary material). We find $a=\unit[4.37]{\AA}$
	and $c=\unit[30.51]{\AA}$ for $\mathrm{Bi_{2}Te_{3}}$
	and $a=\unit[4.45]{\AA}$ and $c=\unit[24.0]{\AA}$
	for $\mathrm{Bi_{1}Te_{1}}$. Besides, the stoichiometries of the
	samples were also checked by Rutherford backscattering spectroscopy
	which confirmed the $50:50$ ratio of $\mathrm{Bi:Te}$ (see section
	I in the supplementary material). \\
	Figure \ref{fig:Bulk-characterization}(b) depicts a high-angular
	annular dark field image of a representative section of a $\unit[39]{nm}$
	thick $\mathrm{Bi_{1}Te_{1}}$ film recorded by STEM. The observed
	clear contrast is related to the difference between individual atomic
	columns of Bi and Te (Bi atomic columns appear brighter than Te columns).
	Distinct van der Waals gaps, separating quintuple layers from Bi bilayers
	are visible and the arrangement of BL and QL matches the expected
	$1:2$ composition ratio. Furthermore, by extracting a line profile
	(yellow frame) and fitting Gaussians to the peaks (green = Te; red
	= Bi) the atomic positions can be determined precisely. Using this
	method, the size of the bulk unit cell was confirmed to be $c=\unit[24.0\pm0.1]{\AA}$.
	\\
	As we have seen, due to the superlattice character of $\mathrm{Bi_{1}Te_{1}}$,
	there is more than one possible surface termination, and neither XRD
	nor STEM probe the surface. Section II in the supplementary information
	presents a spectroscopic study on the chemical composition of the
	surface of $\mathrm{Bi_{1}Te_{1}}$ and the influence of noble gas
	sputtering on the surface termination. It turns out that our growth
	conditions result in Bi-poor surfaces, while ion sputtering leads
	to Bi-rich surfaces. \\
	However, since we expect differently terminated surface terraces to
	be in the order of few micrometer in size \cite{Gibson2013,Valla2012},
	and we employ beam spot sizes of $\unit[400]{\mu m}$ (HR
	ARPES) or even $\unit[1]{mm}$ (spin-ARPES) in the ARPES experiments,
	we cannot exclude that our electronic structure investigations may
	always probe a superposition of different terminations. Therefore,
	due to the rich variety of surface-related states (Fig. \ref{fig:surface DFT})
	and the fact that our ARPES technique lacks lateral resolution, a
	detailed distinction of the surface electronic features may be complicated.\\

\subsection{Surface electronic structure by ARPES. }

\begin{figure*}
	\centering
	\includegraphics[scale=0.9]{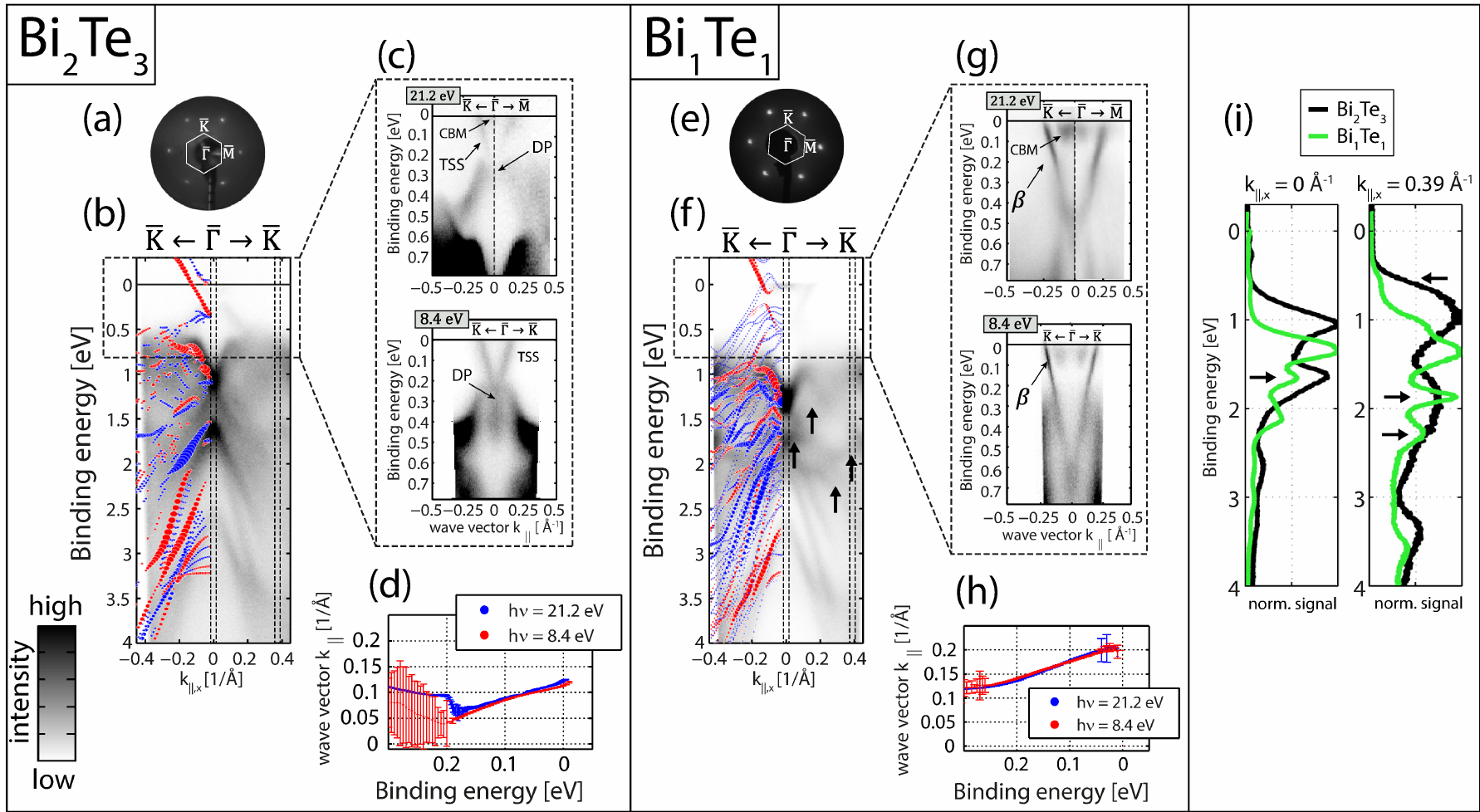}
	\caption{\label{fig:HR ARPES}Comparative ARPES investigation on thin films
	of $\mathrm{Bi_{2}Te_{3}}$ and $\mathrm{Bi_{1}Te_{1}}$ at low temperatures
	$T\approx\unit[25]{K}$. (a) [e] Low-energy electron diffraction
	pattern of a $\unit[19]{nm}$ $\mathrm{Bi_{2}Te_{3}}$ {[}$\unit[45]{nm}$
	$\mathrm{Bi_{1}Te_{1}}${]} film for identification of the orientation
	of the surface Brillouin zone. (b) [f] Wide energy range $E_{\mathrm{B}}$
	vs. $k_{||,x}$ spectra along $\overline{\Gamma\mathrm{K}}$ direction
	of both samples measured with $h\nu=\unit[21.2]{eV}$ with superimposed
	spin-polarized DFT calculations from Fig. \ref{fig:surface DFT} (red
	and blue dots are opposite in-plane spin channels; 1QL termination
	is used for $\mathrm{Bi_{1}Te_{1}}$). Black arrows in (f) mark prominent
	changes compared to (b). Dashed lines mark cuts of the energy distribution
	curves (EDCs) shown in (i). (c) [g] Magnified electronic structure
	close to the Fermi level for two different photon energies $h\nu=\unit[21.2]{eV}$
	and $\unit[8.4]{eV}$ along indicated crystallographic directions.
	Conduction band minimum (CBM), Dirac point (DP) and topological surface
	state (TSS) marked by arrows. (d) [h] Plot of the inverse energy
	dispersion $k(E_{\mathrm{B}})$ of the right branch of the prominent
	TSS {[}$\beta$ state{]} for both photon energies as determined by
	Voigt fits to the momentum distribution curves in (c) [g]. (i)
	EDCs obtained from the spectra shown in (b) (black curve) and (f)
	(green curve) along $k_{||,x}=\unit[0]{\r{A}^{-1}}$ (left)
	and $\unit[0.39]{\r{A}^{-1}}$ (right). Black arrows mark spectral changes corresponding to the arrows in (f). }
\end{figure*}

The comparative results of our ARPES investigations on vacuum transferred,
as-grown $\mathrm{Bi_{2}Te_{3}}$ and $\mathrm{Bi_{1}Te_{1}}$ thin
films is summarized in Fig. \ref{fig:HR ARPES}. As mentioned and
confirmed above, we consider the as-grown $\mathrm{Bi_{1}Te_{1}}$
surface as Bi-poor. In the case of the prototypical STI $\mathrm{Bi_{2}Te_{3}}$
our results nicely reproduce earlier findings \cite{Herdt2013}. \\
In general, the spectra exhibit sharp features and a very good signal
to noise ratio revealing the high crystalline quality of the thin
films. This is also reflected in the low-energy electron diffraction
pattern in Fig. \ref{fig:HR ARPES}(a) and (e), where the orientation
of the surface Brillouin zones is illustrated. Figure \ref{fig:HR ARPES}(b)
and (f) depict wide range binding energy $E_{\mathrm{B}}$ vs. wavevector
$k_{||,x}$ maps of $\mathrm{Bi_{2}Te_{3}}$ and $\mathrm{Bi_{1}Te_{1}}$,
respectively, along trajectories in the $\overline{\Gamma\mbox{K}}$
direction which traverse the $\overline{\Gamma}$ point of the surface
BZ recorded with $h\nu=\unit[21.2]{eV}$. Already on first glance
the spectra of the two samples show a lot of similarities but also
some major differences (e.g. marked by the black arrows). Both samples
are of n-type nature with the conduction band minimum (CBM) being
cut by the Fermi level, but in $\mathrm{Bi_{1}Te_{1}}$ there seems
to be an even stronger considerable downshift of the entire valence
band, due to a possible electron donation of the Bi bilayers to the
QLs \cite{Valla2012}. The downshift as well as the spectral changes
are clearly revealed by the energy distribution curves (EDCs) in (i),
which are obtained along normal emission, i.e., $k_{||,x}=\unit[0]{\AA^{-1}}$
(left) and $k_{||,x}=\unit[0.39]{\text{\AA}^{-1}}$(right), marked
by the dashed area in (b) and (f). Here, black arrows mark the spectral
changes that can be attributed to additional features appearing in
(f). \\
Additionally, in the wide range ARPES maps the spin-polarized surface
electronic structure slab-calculations are superimposed (for 1QL surface
termination in the case of $\mathrm{Bi_{1}Te_{1}}$). Here, red and
blue dots mark oppositely oriented in-plane spin channels and the
size of the dots corresponds to the spin-polarization. The Fermi level
in the calculation needed to be shifted upwards by $\unit[250]{meV}$
{[}$\unit[100]{meV}${]} to fit better to the experimental data of
$\mathrm{Bi_{2}Te_{3}}$ {[}$\mathrm{Bi_{1}Te_{1}}${]}. The prominent
and intense Rashba-type surface state located between $E_{\mathrm{B}}=\unit[0.7-1.05]{eV}$
{[}$\unit[0.95-1.3]{eV}${]} has been used as a gauge to match the
calculation to the ARPES data. As one can see, the agreement between
the data and DFT simulation is very high and most of the features
can be matched. Only the predicted gap-opening along $\overline{\Gamma\mathrm{K}}$
in the uppermost prominent band, which we labeled $\beta$, is not
reproduced in the experimental spectrum. \\
The reason why we do not resolve a gap-opening in the ARPES data,
neither along $\overline{\Gamma\mathrm{M}}$ nor $\overline{\Gamma\mathrm{K}}$
direction, is again most probably due to the lack of lateral resolution
of our measurement technique. The vast variety of surface bands originating
from different terminated areas on the sample may easily provide states
that overlap or even hybridize and thus close the gap. The right panel
in Fig. \ref{fig:surface DFT} shows that the 2QL terminated surface
provides states that close the gap also along $\overline{\Gamma\mathrm{K}}$
direction. Section III in the supplementary material shows how the
superposition of the calculated spectra for both 1QL and 2QL terminated
surfaces hinders the observation of the band gap.\\

Panels (c) and (g) of Fig. \ref{fig:HR ARPES} depict magnified close-Fermi
level spectra of the $h\nu=\unit[21.2]{eV}$ excitation along both
$\overline{\Gamma\mathrm{M}}$ and $\overline{\Gamma\mathrm{K}}$
direction and the same spectra obtained using $h\nu=\unit[8.4]{eV}$
along $\overline{\Gamma\mathrm{K}}$. The two different photon energies
are used to probe a different cut in the 3D Brillouin zone, i.e.,
a different $k_{\perp}$, and to thus provide additional evidence
of the surface state character of the states. Indeed, for $\mathrm{Bi_{2}Te_{3}}$
the TSS, driven by time-reversal symmetry, which is well-known in
literature, is revealed and the Dirac point (DP) is located around
$E_{\mathrm{B}}\approx300$ meV and buried in bulk valence band pockets.
\\
On the other hand, the prominent and interesting $\beta$ feature
in $\mathrm{Bi_{1}Te_{1}}$ seems to disperse strongly linearly and
could, on the first glance, be confused with a topologically non-trivial
Dirac cone state. Indeed, the lack of $k_{\perp}$-dispersion of the
TSS in $\mathrm{Bi_{2}Te_{3}}$ and the $\beta$ band in $\mathrm{Bi_{1}Te_{1}}$
is quantified in Fig. \ref{fig:HR ARPES}(d) and (h), where the wave
vector $k_{||}$ of the right branch of the TSS and the $\beta$ state
is plotted against the binding energy for the two different photon
energies. The data points were extracted out of Voigt peak fits to
the momentum distribution curves of both spectra in Fig. \ref{fig:HR ARPES}(c)
and (g). For the first $\unit[200]{meV}$ below $E_{\mathrm{F}}$,
the fit is very good, i.e., the error very small, but the situation
declines when the states start to hybridize with other bands at higher
energies. The fact that the dispersion of those states is exactly
the same for both $21.2$ eV and $8.4$ eV is a strong indication
of their surface state character. Moreover, from this the Fermi velocity
$v_{\mathrm{F}}$ can be determined by a linear fits as $v_{\mathrm{F}}=\frac{E}{k_{||}\cdot\hbar}$
to be $v_{\mathrm{F}}\approx\unit[2.4]{eV\r{A}}=\unit[3.6\cdot10^{5}]{\frac{m}{s}}$
for $\mathrm{Bi_{1}Te_{1}}$ and $v_{\mathrm{F}}\approx\unit[3.2]{eV\r{A}}=\unit[4.8\cdot10^{5}]{\frac{m}{s}}$
for $\mathrm{Bi_{2}Te_{3}}$.\\
 
\begin{figure}
	\includegraphics[width=0.5\columnwidth]{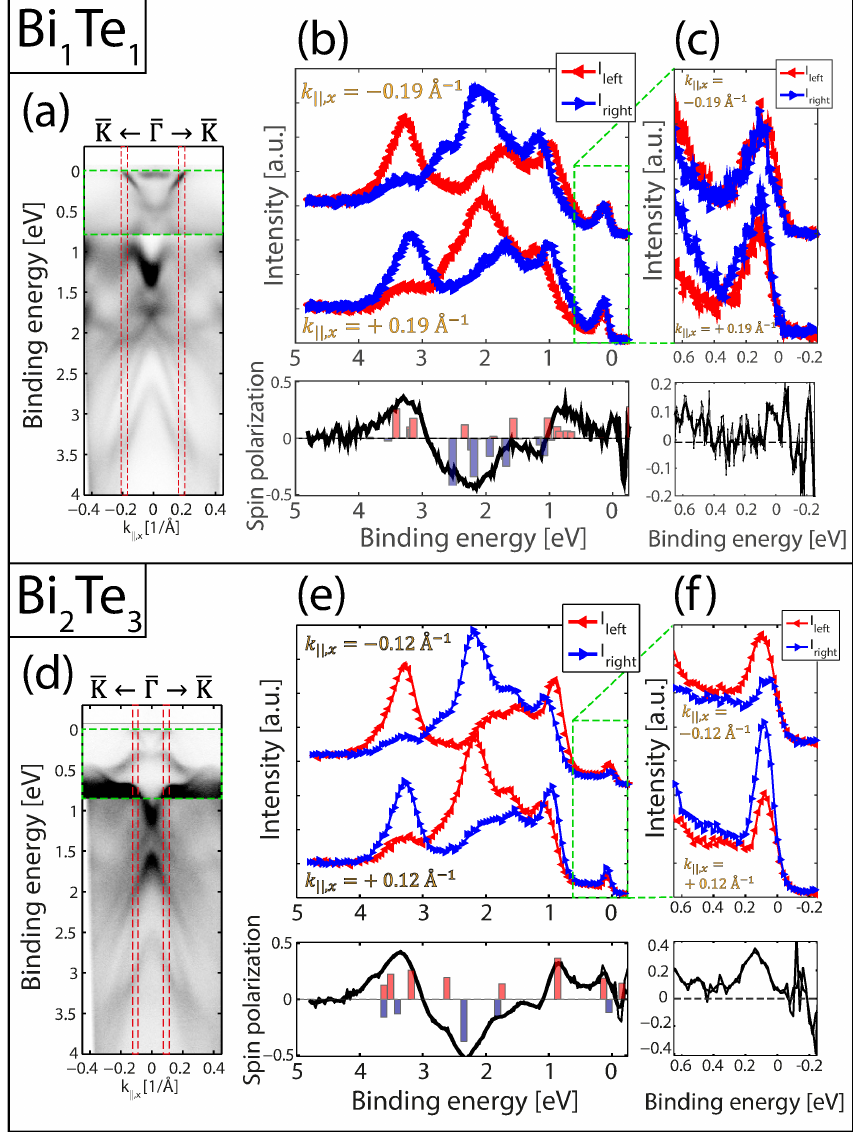}
	\caption{\label{fig:spinARPES}Spin-ARPES investigations recorded with $h\nu=\unit[22]{eV}$
	and a four-channel SPLEED polarimeter. (a) [d] Wide range ARPES
	spectra of $\mathrm{Bi_{1}Te_{1}}$ {[}$\mathrm{Bi_{2}Te_{3}}${]}
	as shown in Fig. \ref{fig:HR ARPES}(b) [f] for the illustration
	of the $k$-points $\left(k_{||,x}=\pm\unit[0.19]{\r{A}^{-1}}\,\left[k_{||,x}=\pm\unit[0.12]{\r{A}^{-1}}\right],\mbox{ red dashed lines}\right)$
	along which the in-plane spin polarization was measured. Note that
	the first $\unit[800]{meV}$ below the Fermi level were boosted in
	contrast (green dashed area). (b) [e] Wide energy spin-polarized
	EDCs at the indicated two opposing $k$-points and effective net spin
	polarization below. The bar graph in the latter shows the calculated
	spin polarization. (c) [f] Magnified EDCs and spin polarization
	close to the Fermi level according to area marked in green in (a) [d]. }
\end{figure}
A strong experimental evidence of the topological nature of a state
is the verification of its helical spin polarization \cite{Ando2013}.
Thus, Fig. \ref{fig:spinARPES} summarizes our findings on the spin
polarization of the $\beta$ state of 'sputtered', i.e., Bi-rich,
$\mathrm{Bi_{1}Te_{1}}$ (a)-(c). Again the data from $\mathrm{Bi_{1}Te_{1}}$
is compared to measurements on $\mathrm{Bi_{2}Te_{3}}$ and the spin
polarization of the prototypical TSS (d)-(f). \\
Figure \ref{fig:spinARPES}(a) and (d) again show the wide range ARPES
maps of $\mathrm{Bi_{1}Te_{1}}$ and $\mathrm{Bi_{2}Te_{3}}$ from
in Fig. \ref{fig:HR ARPES}(b) {[}f{]}, respectively, which illustrate
along which opposing $k$-points, marked by the red dashed area, the
spin polarization is measured. Figures \ref{fig:spinARPES}(b) and
(c) as well as (e) and (f) depict the wide range and close-Fermi level
(in-plane) spin-resolved partial intensities $I_{left}$ and $I_{right}$
along the indicated $k$-points. The spectra were corrected by the
asymmetry function of $S=0.27$, and the net spin polarization is
shown underneath. \\
Both samples show quite similar and rather high in-plane spin polarization
of 40-50\% in the bands at higher binding energies, around $E_{\mathrm{B}}\approx\unit[3.2]{eV}$,
$E_{\mathrm{B}}\approx\unit[2.1]{eV}$ and $E_{\mathrm{B}}\approx\unit[0.9-1.1]{eV}$
in panels (b) and (e). The full reversal of the spin polarization
between the two opposing $k$-points confirms the helical nature of
these states in both samples. Further, the TSS of $\mathrm{Bi_{2}Te_{3}}$
shows a helical spin polarization of up to 40\% in panel (f), which
nicely confirms its topological nature and is in agreement with what
was reported earlier \cite{Herdt2013}.\\
On the contrary, panel (c) reveals that the most interesting $\beta$
state in $\mathrm{Bi_{1}Te_{1}}$ at the Fermi level is measured to
exhibit only very little (though non-vanishing) in-plane spin polarization
of max. 10\% and without a clear reversal at the opposing $k$-points.
Such weak spin polarization can be induced by SOC in topologically
trivial surface states, as most of the states in the calculated band
structures in Fig. \ref{fig:surface DFT} already showed some non-zero
spin polarization. Therefore, this measurement reveals a difference
to the prototypical TSS and thus gives an experimental indication
but no final proof about the topological character of the $\beta$
state. \\

\begin{figure*}
	\centering
	\includegraphics[scale=0.8]{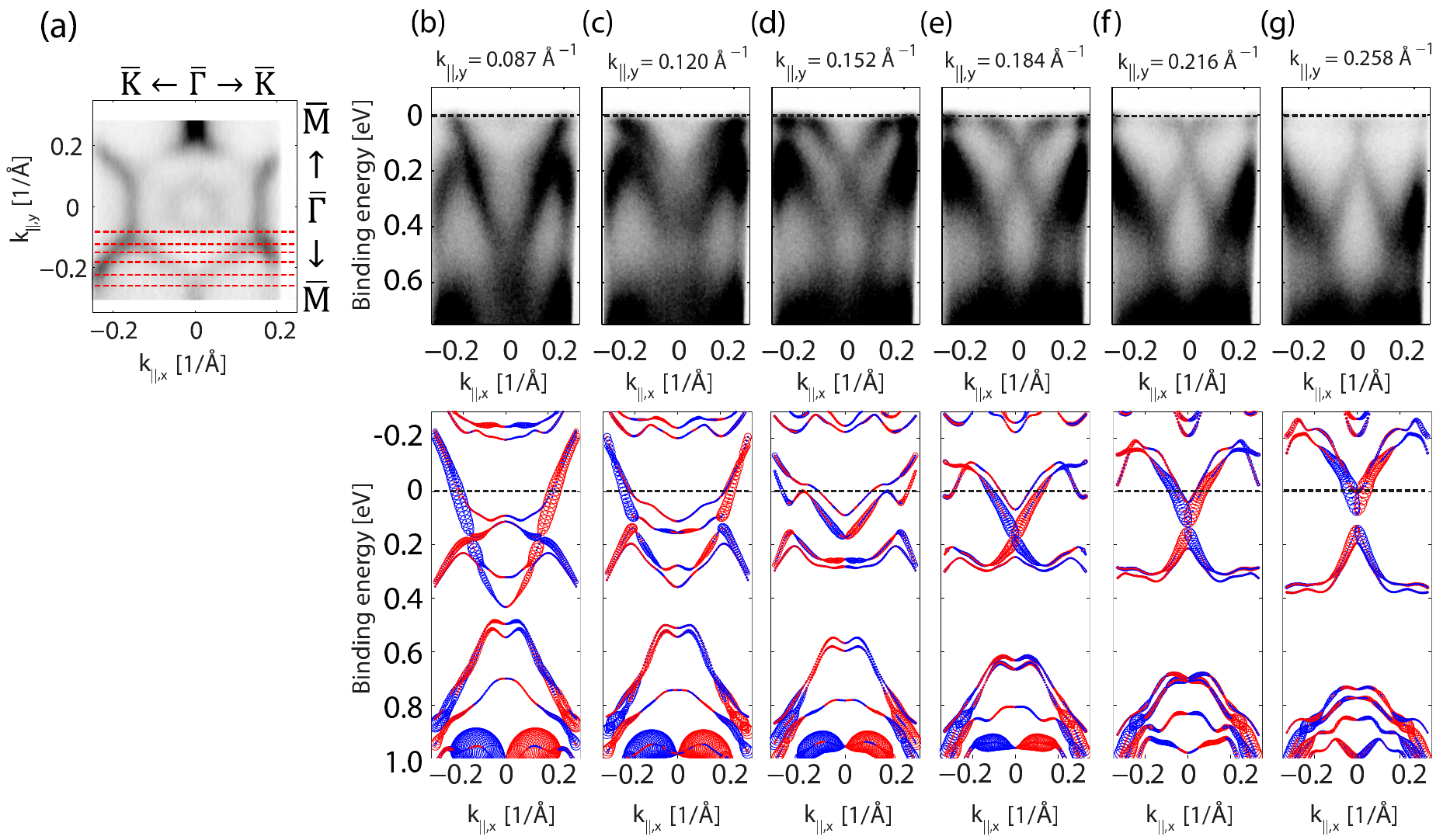}
	\caption{\label{fig:TCI_of normal}Electronic structure of $\mathrm{Bi_{1}Te_{1}}$
	along non-high symmetry lines. (a) Fermi surface $k_{||,x}$ vs. $k_{||,y}$
	map of a $\unit[45]{nm}$ in situ transferred film of $\mathrm{Bi_{1}Te_{1}}$
	obtained with $h\nu=\unit[21.2]{eV}$. Red dashed lines mark trajectories
	along which $E_{\mathrm{B}}(k_{||,x})$ spectra were taken. (b)-(g)
	depict the corresponding spectra as well as spin polarized DFT calculations
	of a 24QL slab of $\mathrm{Bi_{1}Te_{1}}$ with
	1QL terminated surface at $k_{||,y}=\unit[0.087]{\r{A}^{-1}}$(b),
	$\unit[0.120]{\r{A}^{-1}}$(c), $\unit[0.152]{\r{A}^{-1}}$(d),
	$\unit[0.184]{\r{A}^{-1}}$(e), $\unit[0.216]{\r{A}^{-1}}$(f),
	and $\unit[0.258]{\r{A}^{-1}}$(g), respectively. }
\end{figure*}

Finally, we are searching for an experimental evidence of the above
mentioned mirror-symmetry protected band crossings at non-TRIM points,
which are a consequence of the topological crystalline character of
$\mathrm{Bi_{1}Te_{1}}$. Therefore, Fig. \ref{fig:TCI_of normal}
depicts experimental and calculated spectra along non-high symmetry
lines which reveal a region in $k$-space, where the TCI-induced states
can be identified and no other states interfere. Figure \ref{fig:TCI_of normal}(a)
shows a constant energy contour, i.e., $k_{||,x}$ vs. $k_{||,y}$
map, at $E_{\mathrm{B}}=E_{\mathrm{F}}$ with red dashed lines marking
spectra taken along different cut directions at $k_{||,y}=\unit[0.087]{\r{A}^{-1}}$(b),
$\unit[0.120]{\r{A}^{-1}}$(c), $\unit[0.152]{\r{A}^{-1}}$(d),
$\unit[0.184]{\r{A}^{-1}}$(e), $\unit[0.216]{\r{A}^{-1}}$(f),
and $\unit[0.258]{\r{A}^{-1}}$(g). The panels (b)-(g) depict
the respective ARPES spectra obtained with $\unit[8.4]{eV}$ (top)
as well as spin-polarized surface electronic structure calculations
(bottom), which are performed for the 1QL terminated surface. Besides
some additional features from different terminations in the experimental
spectra at higher binding energies, the agreement is very good and
one can clearly identify the interesting bands, which exhibit a mirror-symmetry
protected crossing point around $k_{||,y}=\unit[0.184]{\r{A}^{-1}}$.
We attribute these crossing bands at a non-TRIM point to be the consequence
of the TCI character of $\mathrm{Bi_{1}Te_{1}}$ and to be independent
of the surface termination.

\section{Summary and Conclusions}

In summary, we predicted and demonstrated the dual topological insulator
character of the stoichiometric natural superlattice phase $\mathrm{Bi_{1}Te_{1}}$
by a combined theoretical and experimental investigation. \\
Our study theoretically predicts by \textit{ab initio} DFT calculations
that $\mathrm{Bi_{1}Te_{1}}$ exhibits a 'dark' surface perpendicular
to the stacking direction which is free of time-reversal symmetry
protected surface states at the conventional TRIM points, due to weak
topological indices $\mathbb{Z}_{2}=(0;001)$.
Moreover, we identify an additional protection of topological states
with crossing points at non-TRIM points in the $\overline{\Gamma\mathrm{M}}$
mirror plane direction governed by the crystal mirror-symmetry due
to the non-zero mirror Chern number $n_{{\cal M}}=-2$. This dual
WTI and TCI character of time-reversal and mirror symmetry leads to
the existence of topologically protected states on every surface of
the crystal. \\
Confronting the theoretical predictions with the experiment, we successfully
realized the thin film synthesis of $\mathrm{Bi_{1}Te_{1}}$ on Si(111)
by MBE-growth, carefully characterized the bulk crystal structure
as well as the surface chemistry, and thoroughly investigated the
electronic (spin-) structure. Indeed, we identified significant differences
to the prototypical STI $\mathrm{Bi_{2}Te_{3}}$ in the spin structure
of the surface-related features at the Fermi level, which is a good
indication of the WTI nature of $\mathrm{Bi_{1}Te_{1}}$. Furthermore,
we unambiguously reveal mirror-symmetry protected band crossings at
non-TRIM points which are in excellent agreement to theory and which
we attribute to result from the TCI character of $\mathrm{Bi_{1}Te_{1}}$.
\\
In future work, the weak topological nature of $\mathrm{Bi_{1}Te_{1}}$
could be confirmed by probing the topologically protected one-dimensional
electron edge channels at step edges of the dark surface, e.g. in
a STS study (similar to what was reported in \cite{Pauly2015,Li2016,Wu2016}).\\
The dual topological character opens up new vistas for such materials
in spintronics because topological states are protected by different
symmetries and can be potentially switched on and off by breaking
of a certain symmetry.

\section{Methods}

\subsection{Sample growth.}

All samples for this study are grown as thin films via molecular beam
epitaxy. Firstly, $\unit[10\times10]{mm^{2}}$ Si(111) samples were
prepared by a RCA-HF procedure to remove organic contaminations and
the native oxide. A consecutive HF dip passivates the Si surfaces
with hydrogen for the transfer into the MBE chamber (base pressure
$\unit[5\cdot10^{-10}]{mbar}$). To desorb the hydrogen from the surface
the samples were heated up to $700°C$ for $\unit[10]{min}$
and finally cooled down to 275°C. For the evaporation of
Te and Bi, standard effusion cells were heated to $T_{\mathrm{Te}}=260°C$
and $T_{\mathrm{Bi}}=460°C$, resulting in a growth
velocity of $\mathrm{Bi_{1}Te_{1}}$ of $v=\unit[2.5]{nm/h}$. The
tellurium shutter was opened several seconds in advance to terminate
the silicon surface by Te, which saturates the dangling bonds. While
$\mathrm{Bi_{2}Te_{3}}$ is grown in a tellurium overpressure regime
\cite{Mussler2012}, $\mathrm{Bi_{1}Te_{1}}$ requires equal vapor
pressures of tellurium and bismuth. The 1:1 ratio between bismuth
and tellurium changes the structure from solely quintuple layers in
$\mathrm{Bi_{2}Te_{3}}$ to a layered structure with additional Bi
bilayers between every two QLs in $\mathrm{Bi_{1}Te_{1}}$. \\
After growth, the samples were transferred from the MBE chamber into
the ARPES apparatus ($<\unit[1\cdot10^{-10}]{mbar}$) without breaking
the vacuum by an UHV shuttle with a base pressure below $\unit[1\cdot10^{\text{-9}}]{mbar}$.
The surface of such 'as-grown' samples is, due to the growth mode,
expected to be Bi-poor, i.e., mostly QL-terminated. Nevertheless,
the surface exhibit all three different terminations (see section
II in the supplementary information).\\

\subsection{Structural Characterization.}
	For characterizing the bulk crystal structure, XRD measurements were
	carried out,\textit{ }employing a high-resolution Bruker D8 diffractometer.
	Additionally, crosssectional specimen were measured in an aberration-corrected
	STEM with an electron beam of $\unit[0.8]{\r{A}}$ (FEI Titan
	80-200) for structural investigations on the atomic scale. For this,
	selected specimen are prepared by focused ion beam etching with firstly
	$\unit[30]{keV}$ and subsequently $\unit[5]{keV}$ Ga ions. Later
	Ar ion milling using the Fishione NanoMill was performed to reduce
	the FIB-induced damage. High-resolution STEM images made in high-angular
	annular dark field contain chemical information, since the contrast
	scales with the atomic number $Z^{2}$, allowing to distinguish between
	Bi and Te atoms. \\

\subsection{Spectroscopy.}
	The lab-based high-resolution ARPES investigation was performed at
	$T=\unit[25]{K}$ with a MBS A1 electron spectrometer, using either
	non-monochromatized He I$\mbox{\ensuremath{\alpha}}$ radiation of
	$h\nu=\unit[21.2]{eV}$ from a focused HIS 13 helium lamp or light
	from a microwave-driven MBS xenon discharge lamp producing $h\nu=\unit[8.4]{eV}$
	photons. The beam spot size is about $\unit[400]{\mu m}$ in the former and
	1 mm in the latter case and the light is unpolarized. The analyzer
	measures $E_{\mathrm{B}}$ vs $k_{||,x}$ dispersion maps at once.
	Fermi surface mapping is achieved by rotating the sample with respect
	to the entrance slit of the spectrometer. The overall energy resolution
	is estimated to be $\unit[10]{meV}$ and the angular resolution is $<\unit[0.02]{\r{A}^{-1}}$.
	\\
	For spin-resolved ARPES measurements we used photons of $h\nu=\unit[22]{eV}$
	a Scienta SES-2002 spectrometer and a Focus SPLEED polarimeter at
	beamline BL5 of the DELTA synchrotron in Dortmund at room temperature,
	resulting in an energy resolution of $\approx\unit[100]{meV}$ \cite{Plucinski2010a}.
	Here, clean sample surfaces are prepared by sputtering and annealing
	after sample transfer through air, which resulted in Bi-rich sample
	surfaces (see section II in the supplementary information).\\

\subsection{Electronic structure calculations.}

The DFT calculations are performed for the bulk phase and thin films
with three different surface terminations, namely a single and a double
QL (1 and 2QL), and a Bi BL. The bulk unit cell consists of two QLs
and one Bi BL, and the hexagonal atomic planes are all assumed to
have a fcc-like (A-B-C) stacking. To simulate a Bi BL terminated surface,
a symmetric 26 layer film with BL-QL-QL-BL-QL-QL-BL stacking was used.
For the 1 and 2QL termination, symmetric 24 and 34 layer films were
set up. We employ the full-potential linearized augmented plane wave
method as implemented in the \textsc{Fleur} code \cite{flapw} with
the relaxed lattice parameters from the Vienna \textit{ab-initio}
simulation package \cite{Kresse1996,Kresse1996b}. The generalized
gradient approximation of Perdew-Burke-Ernzerhof form \cite{Perdew1996}
is used for the exchange correlation potential. Spin-orbit coupling
is included self-consistently in the calculations.\\
From the DFT calculations, we obtain structural parameters that are
in good agreement with the experimental data. The size of the bulk
unit cell in $c$-direction is $\unit[25.0]{\r{A}}$. It consists of two QLs of $\unit[7.48]{\r{A}}$ thickness each and
a Bi BL of $\unit[1.68]{\r{A}}$. The BL-QL separation
is $\unit[2.66]{\r{A}}$ and the distance between the QLs
is $\unit[3.04]{\r{A}}$. At the surfaces, these distances
contract slightly, e.g., the QL-QL distance decreases by $\unit[0.06]{\r{A}}$
at the 2QL-terminated surface, while the QL-BL distance is reduced
only by $\unit[0.04]{\r{A}}$ for the 1QL termination.
For BL termination, the interlayer distance changes even less. The
step-height between a BL-terminated and a 2QL-terminated surface is
thus $1.68+2.66=\unit[4.34]{\r{A}}$.

\begin{acknowledgments}
	\section*{Acknowledgements}
The authors acknowledge financial support from the priority program
SPP1666 and the Virtual Institute for Topological Insulators (VITI).
We thank Volkmar Hess, Samuel K\"onigshofen, Frank Matthes, and Daniel
B\"urgler for additional characterization of the surface chemistry of
our samples in their laboratory-based XPS chamber. Also we thank B. Holl\"ander for the precise determination of the samples stoichiometry via RBS. 
Further, the authors acknowledge the technical support by B. K\"upper and A. Bremen. 
\end{acknowledgments}


%

\end{document}